# Aging, Fragility and Reversibility Window in Bulk Alloy Glasses

*S. Chakravarty, D.G. Georgiev, P.Boolchand*

Department of ECECS, University of Cincinnati, Cincinnati, OH 45221-0030,USA

*M.Micoulaut*
Laboratoire de Physique Theorique des Liquides, Universite Pierre et Marie Curie,
Boite 121, 4 Place Jussieu, 75252 Paris, Cedex 05, France

## Abstract

Non-reversing relaxation enthalpies ($\Delta H_{nr}$) at glass transitions $T_g(x)$ in the $P_xGe_xSe_{1-2x}$ ternary display wide, sharp and deep global minima (~ 0) in the $0.09 < x < 0.145$ range, within which $T_g$s become *thermally reversing*. In this *reversibility window*, glasses are found *not to age*, in contrast to *aging* observed for fragile glass compositions outside the *window*. *Thermal reversibility* and *lack of aging* seem to be paradigms of self-organization which molecular glasses share with *protein* structures which repetitively and reversibly change conformation near $T_g$ and the folding temperature respectively.

Aging occurs in many materials, both organic and inorganic. Inorganic crystals age under electrical, mechanical, or thermal stresses, often as a result of dislocation motion. Aging in organic materials is more complex, occurring as hydrogen bonding configurations are altered as a result of thermal cycling. The more complex the system, the less understood are the causes of aging, and the causes of aging in living systems are one of the greatest unsolved mysteries in science. Here we will show that inorganic non-crystalline nanonetworks are universally divided into three regimes of composition, two of which age rapidly, while the third regime scarcely ages at all. The third regime defines a narrow window of composition that appears to have much in common mechanically with selected organic nanonetworks, namely the polypeptide chains that form proteins. Aging is not evident in data obtained by conventional structural methods (diffraction, infrared and Raman spectra, magnetic resonance), but it is measured very accurately by thermally modulated scanning calorimetry.

New ideas on the nature of glass transitions ($T_g$) have emerged in recent years from examination[1-3] of the *non-reversing relaxation enthalpy* ($\Delta H_{nr}$) associated with $T_g$. The endotherm is signature of ergodicity breaking events when structural arrest of a glass forming liquid occurs near $T_g$. Examined as a function of mean coordination number, *r*, of network glasses, the endotherm ($\Delta H_{nr}$) is found to nearly vanish[1-5] across *compositional windows*, $r_c(1) < r < r_c(2)$, wherein glass transitions become *thermally reversing*. Furthermore, these *reversibility windows* are found to be closely related to variations in Raman optical elasticities[3-5]. Distinct elastic power-laws are seen in the composition regimes: $r < r_c(1)$), in $r_c(1) < r < r_c(2)$ (the window), and above $r > r_c(2)$ [3-5]. Theoretical support for the existence of these elastic phases comes from several related considerations, namely, the counting of Lagrangian constraints[6], graph theory[7] and numerical simulations[7,8]. From these considerations J.C.Phillips and M.F. Thorpe have identified[6-8] the existence of *three* generic elastic phases as a function of *r*; *floppy, intermediate and stressed rigid*. Thus, $r_c(1)$ and $r_c(2)$, mark the *onset* and *end* of the *reversibility window*, and are the two *phase boundaries* between these *three elastic phases*. Intermediate phases are generally centered on mean coordination numbers close to 2.40, and are thought to consist of isostatic (rigid but unstressed) structures. In the present work we examine the rates of network aging in the three composition regimes as measured by changes in the kinetics of the glass transition in samples relaxed at room temperature (far below the glass transition temperatures) over several months. Our results show that strain plays an important part in network aging, and that the absence of long-range strains is a necessary condition for non-aging.

Intuitively one might anticipate non-aging in the very stable and self-organized structure that could exist between the two less stable structure regimes, namely *floppy* structures with polymer chain entanglements frozen upon quenching and overconstrained *stressed-rigid* structures with locked-in extra bonds on the other. The caveat, however, is the possible presence of *nanoscale phase separation* which invalidates mean field constraint counting particularly in the would be optimally constrained phase. Such separation often occurs in binary selenides, $(T \text{ or } Pn)_x Se_{1-x}$, where *T* is a tathogen (Si, Ge) and *Pn* a pnictide (P,As), and manifests itself in global maxima of $T_g(x)$ near chemical thresholds[10,11] ( Fig. 1a). However, in ternary selenides, $T_x Pn_x Se_{1-x}$, containing *equal* *fractions* of *T* and *Pn* atoms, these global maxima are conspicuously absent[12], and $T_g(x)$ is found to increase monotonically with x, as illustrated for the case of $T$ = Ge, $Pn$ = P ternary in Fig.1a. In this Letter, we identify the r*eversibility window* (0.09 < x < 0.145) in the $P_xGe_xSe_{1-2x}$ ternary and find it to be wide, sharp and deep. Furthermore, glasses *in the window* are found not to age, in sharp contrast to aging observed for glass compositions *outside* the window. *Thermal reversibility and absence of aging* appear to be common features of both *glasses* in *reversibility windows* and *proteins in folding states*[13,14], and are connected with structural *self-organization* of these disordered networks.

Our dry, homogenized samples were prepared as described previously[3]. Glass transition temperatures, $T_g(x)$, and non-reversing relaxation enthalpy, $\Delta H_{nr}(x)$, were established using a model 2920 MDSC from TA instruments. Measurements were performed on fresh (3 weeks) and aged (3 and 5 months) samples relaxed at 300K. We find $T_g(x)$ to increase monotonically with x in the 0 < x < 0.25 range ( Fig.1a). Variations in $\Delta H_{nr}(x)$ show (Fig. 1b) a global minimum in the 0.09 < x < 0.145 range that becomes relatively sharper and deeper as the glasses outside the window *age*. In the 0.20 < x < 0.23 range, variations in $T_g(x)$ and $\Delta H_{nr}(x)$ show a mild glitch ( Fig.1a) and a satellite window (Fig.1b) respectively. Raman scattering on glasses excited in the IR ( 1.06 μm ) were performed in a back scattering geometry using a Nicolet FT Raman module with model 870 FTIR bench at 1 cm$^{-1}$ resolution. Fourteen bands were identified and their strengths traced as functions of composition in order to monitor the nature of the molecular clusters and the degree of nanoscale phase separation. This allows the identification (Fig. 2a) of P- centered[15-18] pyramidal (PYR) and quasi-tetrahedral (QT), and Ge-centered[5,19,20] corner-sharing (CS) and edge-sharing (ES) units from their vibrational modes. The nature of P-centered units is independently confirmed from $P^{31}$ NMR results[21]. These considerations lead to the construction of a full ternary phase diagram showing the regimes of the three generic elastic phases observed near the stiffness transition in these alloys; details will be published elsewhere. Molar volumes of the glasses in the fresh and aged state were measured (Fig.3) using Archimedis method.

The central result of the present work is the observation of a deep and wide *reversibility window* in the 0.09 < x < 0.145 (or 2.27 < *r* < 2.44) range. The window delineates the *three elastic phases*; *floppy* at *r* < 2.27, *intermediate* in the 2.27 < *r* < 2.44 range, and *stressed rigid* at *r* > 2.44. Here *r* = 2 +3x (ref.2). The *reversibility window sharpens* and gets *deeper* as glass compositions outside the *window* age at 300K (Fig. 1b). Floppy glasses (below the window) age over a 3-month waiting period, while *stressed-rigid* glasses (above the window) age somewhat slower ~ 5-months. The slower aging kinetics of the latter phase is probably connected with their higher $T_g$s. There is no evidence of aging for glasses in the *reversibility window* even after a 5 - month waiting period.

In the reversibility window composition range the local structures are thought to consist of CS ( *r* = 2.40-2.67) and ES ( *r* = 2.67) Ge(Se$_{1/2}$)$_4$ tetrahedra, pyramidal ( *r* = 2.40) P(Se$_{1/2}$)$_3$ and quasi-tetrahedral ( *r* = 2.28) Se=P(Se$_{1/2}$)$_3$ units. These structural units all have the feature[2] that the number of Lagrangian constraints/atom due to bond-stretching and bond-bending forces equals 3, the degrees of freedom/atom according to which they comprise *isostatic rigidity*[7,8]. The exceptional thermal and elastic behavior of glasses in the *reversibility window* derives from the isostatically rigid nature of their backbones. From theoretical considerations it is plausible that the backbones of these alloys can be made up entirely of such isostatic units over the range of the *reversibility window*[2,3,5,9,22]. Thus, for example, the window begins near $r = 2.28$ where the concentration of QT unit ($r = 2.28$) maximizes[21] (Fig.2b), and the window ends near $r = 2.44$ where concentrations of PYR units ($r = 2.40$), CS ($r = 2.40$-$2.67$) and ES units ($r = 2.67$) is high (Fig.2a). Furthermore, one would expect molecular packing of these units as manifested in molar volumes of the glasses to show absence of aging effects as is indeed observed (Fig. 3) for window compositions. In the latter, molar volumes are found to be nearly independent of coordination number.

In conclusion, MDSC and Raman scattering on $P_xGe_xSe_{1-2x}$ glasses have permitted isolation of the *reversibility window* and analysis of its average and local structures. One of the popular qualitative descriptions of glasses describes the temperature dependence of the viscosity of a supercooled melt in terms of whether it exhibits a constant Arrhenius activation energy, or whether this energy increases as the melt is supercooled; the former materials are said to form strong glasses, the latter fragile ones[23]. In As-Se binary glasses we have already pointed out that the reversibility window coincides with a window in activation energies for viscosity relaxation[3]. The essential new feature here is the evidence that for the same composition range there does not seem to be structural aging in sharp contrast to the behavior encountered for fragile glass compositions outside this window.

The close correspondence between *reversible melting and aging* near the stiffness transition of the network backbone are most suggestive for the analogy with protein folding which has been postulated in recent skeletal models[13] of living polypeptide chains. There it was shown that 26 diverse proteins undergo a mechanical stiffness transition very similar to that in singly bonded network glasses near the same average coordination number (<**r**> = 2.41) found in the glasses[13]. Noteworthy also are the DSC results[24] on a variety of proteins (large ones such as Lys- plasminogen that have several loops or single domain proteins such as egg-white lysozyme ) that show the denaturation process to be *thermally reversible*. Repetitive calorimetric scans[24] taken across the folding transition(s) in each instance are found to be almost indistinguishable from each other. Of course, protein functionality demands that there be almost no aging and nearly complete reversibility during the life of the protein. It is in this respect the present finding of a *reversibility window* in glasses bears a close analogy to protein folding.

The width of the windows may be determined by the strengths of residual interactions relative to the covalent constraints. The narrow width of the protein window Δ**r** = ±0.02**,** compared to <**r**> = 2.36 and Δ**r** = ±0.08 for the glasses studied here, indicates a much more *sophisticated* degree of self-organization achieved in proteins by evolutionary design. In glasses these residual fluctuations may be the external covalent bonds between P-pyramids and Ge-tetrahedra, while in the proteins the important residual interactions involve weak H bonds[13]. Note that a very narrow window centered at **r** = 2.34 of width Δ**r** = ±0.0025 has been observed in Ge-S-I glasses; this width may reflect the weakness of non-bonded I-I van der Waals interactions[25]. Here we have shown that the conditions for formation of reversible functionality in glasses and proteins are similarly distinctive and can be characterized by the mechanical properties of their elastic backbones. We thank M. Mabry and B.Zuk of ThermoNicolet Inc. for the Raman measurements. LPTL is Unite Mixte de Recherche CNRS No 7600. This work is supported by NSF grant DMR-01-01808.

References

1. D.G.Georgiev, P.Boolchand and M. Micoulaut, Phys. Rev B **62**, R9228 (2000).
2. Y.Wang, P.Boolchand and M.Micoulaut, Europhys. Lett. **52**, 633 (2000).
3. P.Boolchand, D.G.Georgiev and M. Micoulaut, J.Optoelectronic.Adv.Mater. **4**, 823 (2002). Also see ibid **3**,703(2001).
4. D.Selvanathan, W.J.Bresser and P.Boolchand, Phys. Rev. B **61**, 15061 (2000).
5. Tao Qu, D.G.Georgiev, P.Boolchand and M.Micoulaut in *Supercooled Liquids, Glass Transition and Bulk Metallic Glasses*" (eds T. Egami, A. L. Greer, A. Inoue, S. Ranganathan). Mater. Res. Soc. Symp. Proc. vol. **754**, CC8.1.1 (2003)
6. J.C.Phillips, J.Non Cryst. Solids **34**, 153 (1979).
7. M.F.Thorpe, D.J.Jacobs, N.V.Chubynsky and A.J.Rader in *Rigidity Theory and Applications,* Ed. M.F.Thorpe and P.M. Duxbury, Kluwer Academic/Plenum Publishers, 1999, p.239
8. M.F.Thorpe, D.J.Jacobs, M.V.Chubynski and J.C.Phillips, J.Non-Cryst. Solids **266-269**, 859 (2000).
9. J.C.Phillips, Phys. Rev. Lett., **88**,216401 (2002).

10. R.Kerner and M.Micoulaut, J. Non-Cryst. Solids **210**, 298 (1997).

11. M.Micoulaut, Eur.Phys. J., B**1**, 277(1998).

12. P.Boolchand, D.G.Georgiev, T.Qu, F.Wang, L.Cai and S. Chakravarty, Comptes Rendus Chimie **5**, 713 (2002).

13. A.J.Rader, B.M. Hespenheide, L.A.Kuhn and M.F.Thorpe, Proc. Nat. Acad. Sci. USA **99**, 3540 (2002); M. F. Thorpe, APS News (2), 10 (2003).

14. A. Fersht, *Structure and Mechanism in Protein Science: A Guide to Enzyme Catalysis and Protein Folding* ( Freeman, New York, 1999).

15. W. Bues, M.Somer and W.Brockner, Z.Naturforsch., **35b**, 1063 (1980).

16. K.Andreas, K.Alexander and T.Martin, J.Chem.Phys., **116**, 3323 (2002).

17. D.G. Georgiev, M.Mitkova, P.Boolchand, G.Brunklaus, H.Eckert and M.Micoulaut, Phys. Rev. B **64**, 134204 (2001).

18. K.Jackson, A.Briley, S.Grossman, D.V.Porezag and M.R.Pederson, Phys. Rev. B **60**, R14985 (2001).

19. K.Murase, in *Insulating and Semiconducting Glasses*, Ed. P.Boolchand ( World Scientific Press Inc., Singapore, 2000) p. 415

20. Xingwei Feng, W.J. Bresser and P.Boolchand, Phys. Rev. Lett. **78**, 4422 (1997).

21. C. Lyda, T.Tepe, M.Tullius, D.Lathrop and H.Eckert, J.Non-Cryst. Solids **171**, 271 (1994).

22. M. Micoulaut and J.C.Phillips, Phys. Rev. B **67**, 104204 (2003).

23. C. A. Angell , J.Non Cryst. Sol. **102**, 205 (1988). ibid., **131 and 133**, 13 (1991).

24. P.L.Privalov, J. Chem. Thermodynamics **29**, 447(1997).

25. Y. Wang, J. Wells, D. G. Georgiev, P. Boolchand, K. Jackson, and M. Micoulaut, Phys. Rev. Lett. **87**, 185503 (2001); J. C. Phillips (unpublished).

Captions

**Fig.1 (a)**. $T_g(r)$ trends in Ge-Se (◊) , P-Se (□) and P-Ge-Se (●) glasses. The thick black line shows the $T_g(r)$ prediction[10,11] based on SAT. Inset shows concentration of homopolar bonds projected by SAT to account for the observed $T_g(r)$ trend. The x-axis scale for inset is the same as that of the ternary in Fig.1a. **(b)** Trends in $\Delta H_{nr}(x)$ in the $Ge_xP_xSe_{1-2x}$ ternary showing the reversibility window in the $0.09 < x < 0.145$; the latter gets deeper and sharper upon aging of glass samples at 300 K.

**Fig. 2 (a)**. Raman scattering of a ternary glass at x = 0.10 showing modes[15-21] of quasi-tetrahedral (QT) units ( 500 cm$^{-1}$) , ethylenelike (ETH) $P_2Se_3$ units( 375 cm$^{-1}$) , pyramidal (PYR) $P(Se_{1/2})_3$ units ( 330 cm$^{-1}$) , $Se_n$ chain mode (CM) at 250 cm$^{-1}$ and 140 cm$^{-1}$, corner-sharing (CS) and edge-sharing ES $Ge(Se_{1/2})_4$ units near 200 cm$^{-1}$ and 217cm$^{-1}$ respectively(ref 17). **(b)** shows a plot of Raman scattering strength of QT mode normalized to the CM at 250 cm$^{-1}$ in open circles, while filled circles give concentrations of the QT units inferred from $^{31}$P NMR, ref. 21.

**Fig. 3**. Molar volumes of present glasses measured 2 months (●) and 6 months (○) after water quench. Note aging effects occur for glass compositions outside the reversibility window but not inside the window.

Table 1

| Network | Intermediate Phase $r_1, r_2$ | Ref. |
|---|---|---|
| Ge-Se | 2.40, 2.52 | 20 |
| Si-Se | 2.40, 2.53 | 4 |
| As-Se | 2.29, 2.37 | 1 |
| P-Se | 2.28, 2.40 | 17 |
| Ge-S-I | 2.332, 2.342 | 24 |
| Ge-As-Se | 2.27, 2.46 | 5 |
| P-Ge-Se | 2.27, 2.43 | present |
| Proteins | 2.39, 2.42 | 13 |

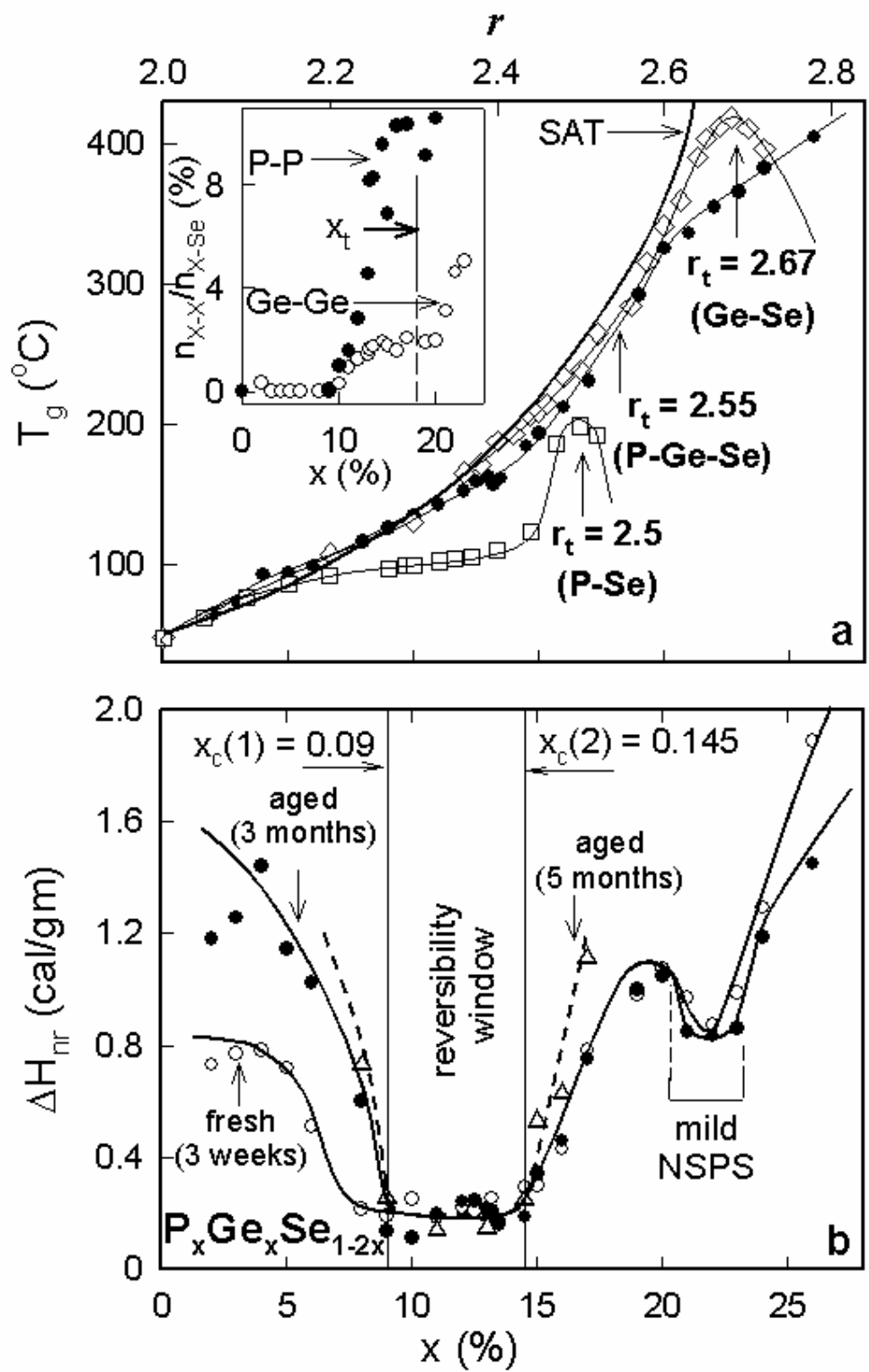

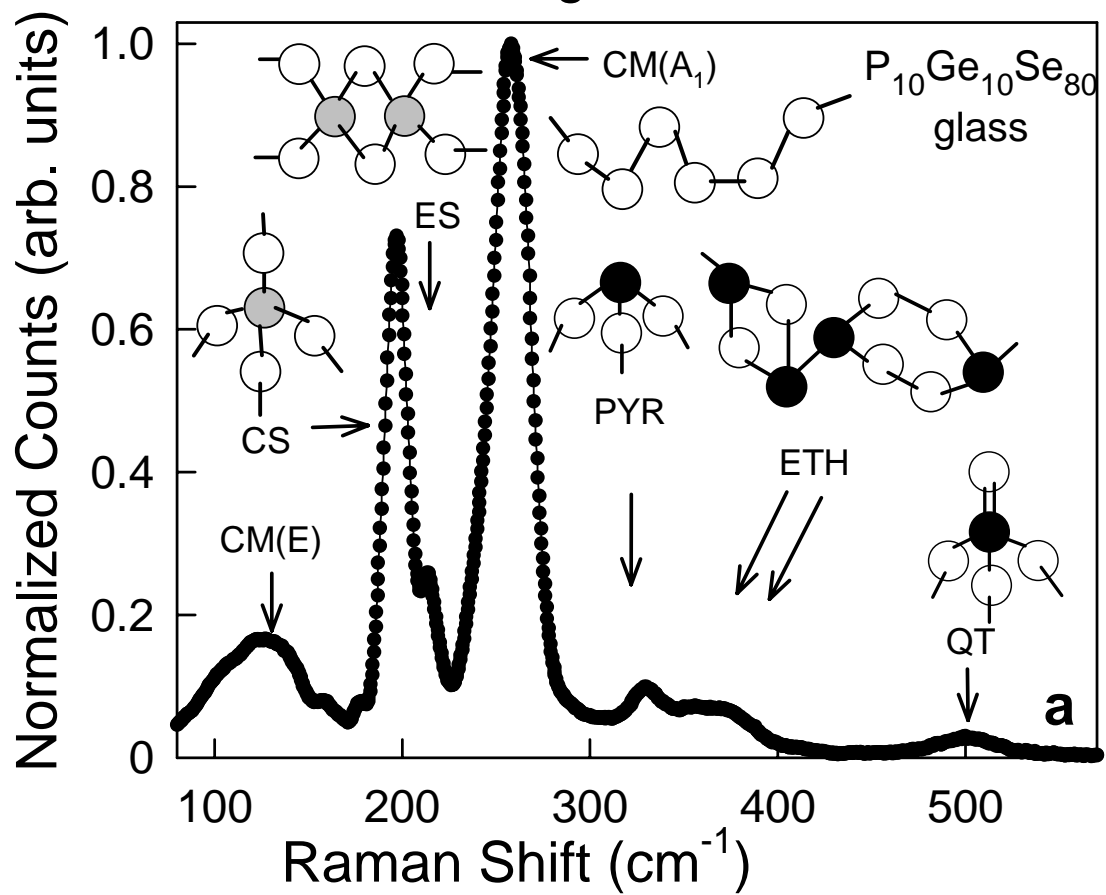

Fig. 2a

Fig. 2b

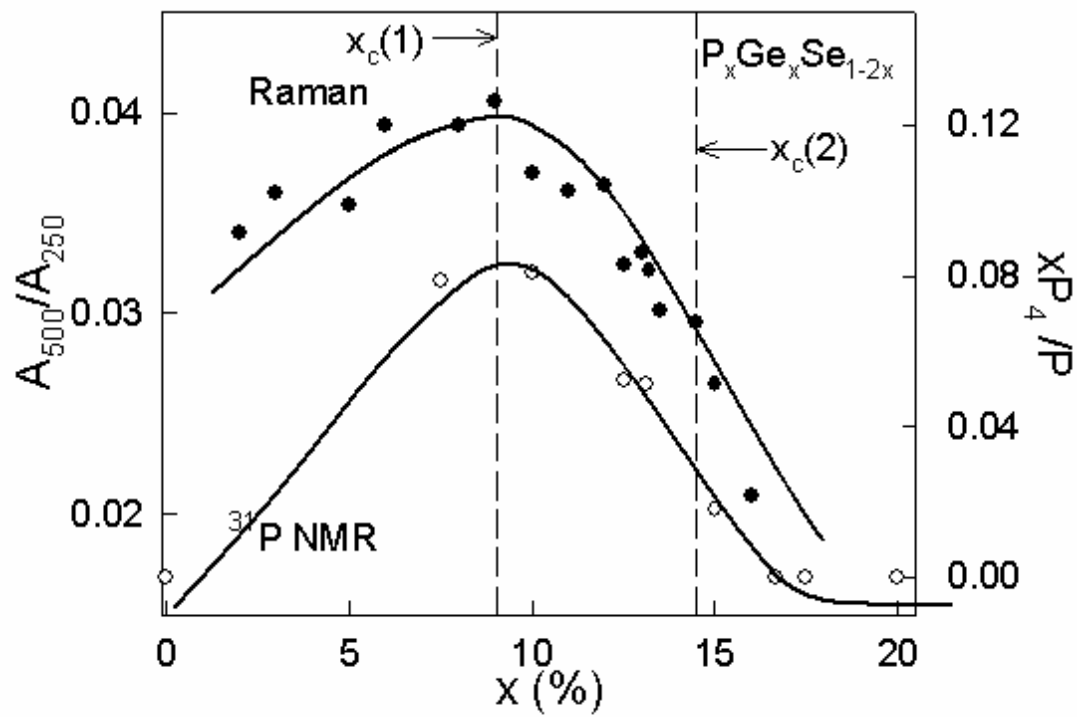

Fig. 3

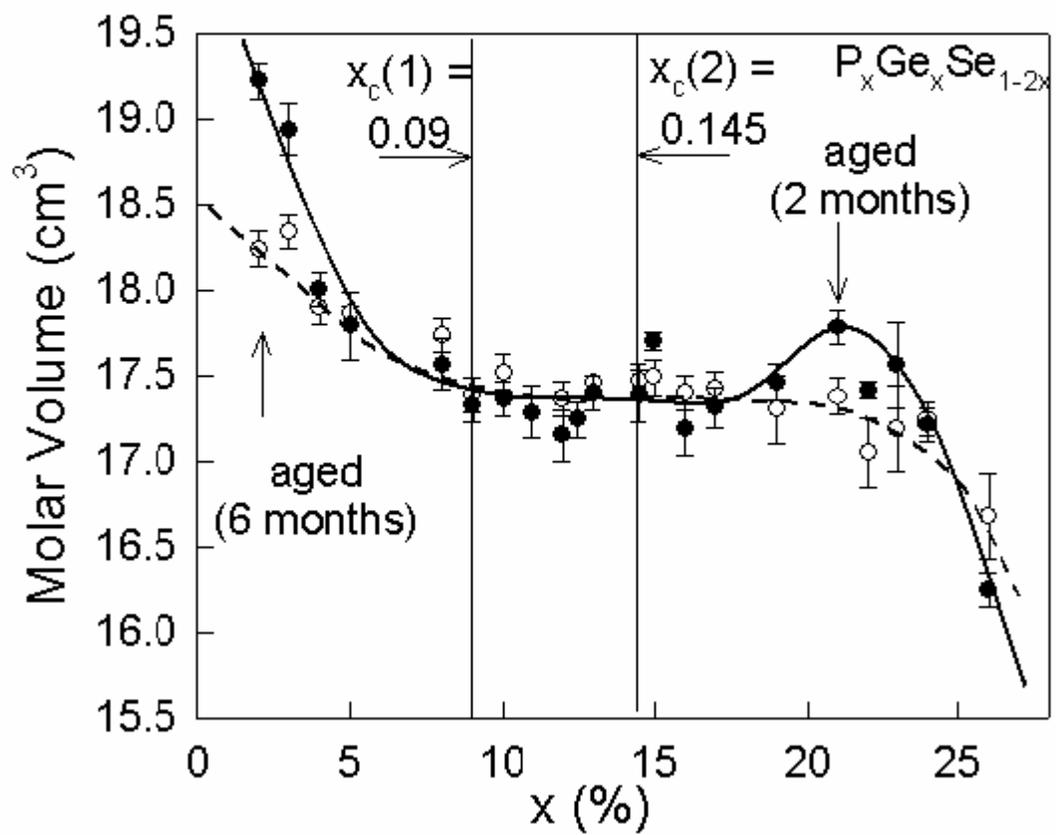